**Room Temperature Initialisation and Readout of Intrinsic Spin Defects in a Van der Waals Crystal**

Andreas Gottscholl[1], Mehran Kianinia[2], Victor Soltamov[1], Carlo Bradac[2], Christian Kasper[1], Klaus Krambrock[3], Andreas Sperlich[1], Milos Toth[2], Igor Aharonovich[2,*], Vladimir Dyakonov[1,*]

[1] Experimental Physics VI and Würzburg-Dresden Cluster of Excellence ct.qmat, Julius Maximilian University of Würzburg, 97074 Würzburg, Germany
[2] School of Mathematics and Physical Sciences, University of Technology Sydney, Ultimo, NSW 2007, Australia
[3] Departamento de Física, Universidade Federal de Minas Gerais (UFMG), Belo Horizonte, MG, Brazil

Correspondence to igor.aharonovich@uts.edu.au or vladimir.dyakonov@uni-wuerzburg.de

**Abstract**
*Optically addressable spins in wide-bandgap semiconductors have become one of the most prominent platforms for exploring fundamental quantum phenomena. While several candidates in 3D crystals including diamond and silicon carbide have been extensively studied, the identification of spin-dependent processes in atomically-thin 2D materials has remained elusive. Although optically accessible spin states in hBN are theoretically predicted, they have not yet been observed experimentally. Here, employing rigorous electron paramagnetic resonance techniques and photoluminescence spectroscopy, we identify fluorescence lines in hexagonal boron nitride associated with a particular defect—the negatively charged boron vacancy ($V_B^-$)—and determine the parameters of its spin Hamiltonian. We show that the defect has a triplet (S = 1) ground state with a zero-field splitting of ≈3.5 GHz and establish that the centre exhibits optically detected magnetic resonance (ODMR) at room temperature. We also demonstrate the spin polarization of this centre under optical pumping, which leads to optically induced population inversion of the spin ground state—a prerequisite for coherent spin-manipulation schemes. Our results constitute a leap forward in establishing two-dimensional hBN as a prime platform for scalable quantum technologies, with extended potential for spin-based quantum information and sensing applications, as our ODMR studies on hBN - NV diamonds hybrid structures show.*

The emergence of 2D materials and van der Waals (vdW) crystals has enabled the observation and realisation of unique optoelectronic and nanophotonic effects such as unconventional superconductivity, Moire excitons and quantum spin Hall effects at elevated temperatures, to name a few.[1-3] Amidst the large variety of studied vdW crystals, hexagonal Boron Nitride (hBN) offers a combination of unique physical, chemical and optical properties[4]. Most relevant to this work, is the ability of hBN to host atomic impurities (or point defects), that give rise to quantized optical transitions, well below its bandgap.[5,6] hBN colour centres are ultrabright with narrow and tuneable linewidth,[7-9] and photostability up to 800 K.[10] Whilst the nature of many of the defects is still uncertain,[11-15] they are being extensively studied as promising candidates for quantum photonic applications requiring on-demand, ultrabright single-photon emission.

A step forward, which will significantly extend the functionality of hBN emitters for quantum applications, is to interface their optical properties with spin transitions, and realise spin-polarization and optical spin-readout schemes.[16, 17] The concept of spin-photon interface has been extensively studied in quantum dots[18] and the nitrogen-vacancy (NV) centre in diamond.[19, 20] The latter has been harnessed to realise basic two-node quantum networks[19] and a plethora of advanced quantum sensing schemes.[21-23] The basic principle is that the triplet spin ground state of the defect can be polarised, manipulated and read out optically owing to the spin-dependent excitation, decay and intersystem crossing pathways available to the system during the optical excitation-recombination cycle.[24]

Yet, extending the optical control of single-spin states beyond defects in 3D crystals, to 2D systems, has remained elusive. If achieved, it will open up a stretch of novel possibilities both fundamental and technological. The two-dimensional nature of these materials inherently allows for seamless integration with heterogeneous, opto-electronic devices where the hosted solid-state qubits can be readily interfaced with cavities, resonators and nanophotonic components from foreign materials. Further, it naturally grants nanoscale proximity of the spin probe to target samples for high-resolution quantum sensing realizations. Reliable and deterministic transfer of hBN layers on stacks of other 2D materials is well-established and is part of one of the currently most relevant endeavours of condensed matter physics—engineering heterostructures made by purposefully-chosen sequences of atomically-thin 2D-materials.[25]

Here we report on the optical initialisation and readout of an ensemble of spins in hBN. We perform rigorous electron paramagnetic resonance (EPR) spectroscopy and optically detected magnetic resonance (ODMR) measurements to establish that the defect has a triplet ground state with zero field splitting (ZFS) of $D/h \approx -3.5$ GHz and almost isotropic Landé factor $g = 2.000$. From the analysis of the angular dependence and nitrogen hyperfine structure, we confirm the intrinsic nature of the defect and assign it to the negatively-charged boron vacancy ($V_B^-$). The alternative nitrogen vacancy ($V_N^0$) structure was also considered, but discarded upon analysis of the experimental data (see discussion below).

Figure 1a is a schematic illustration of the proposed defect. The defect is a negatively charged boron vacancy ($V_B^-$) centre consisting of a missing boron atom surrounded by three equivalent nitrogen atoms in the hBN lattice. The defect has $D_{3h}$ point-group symmetry, typical for a substitutional defect in hBN, and exhibits a strong room temperature photoluminescence (PL) emission at $\lambda_{max} \approx 850$ nm, under $\lambda_{exc} = 532$ nm laser excitation (Figure 1b).

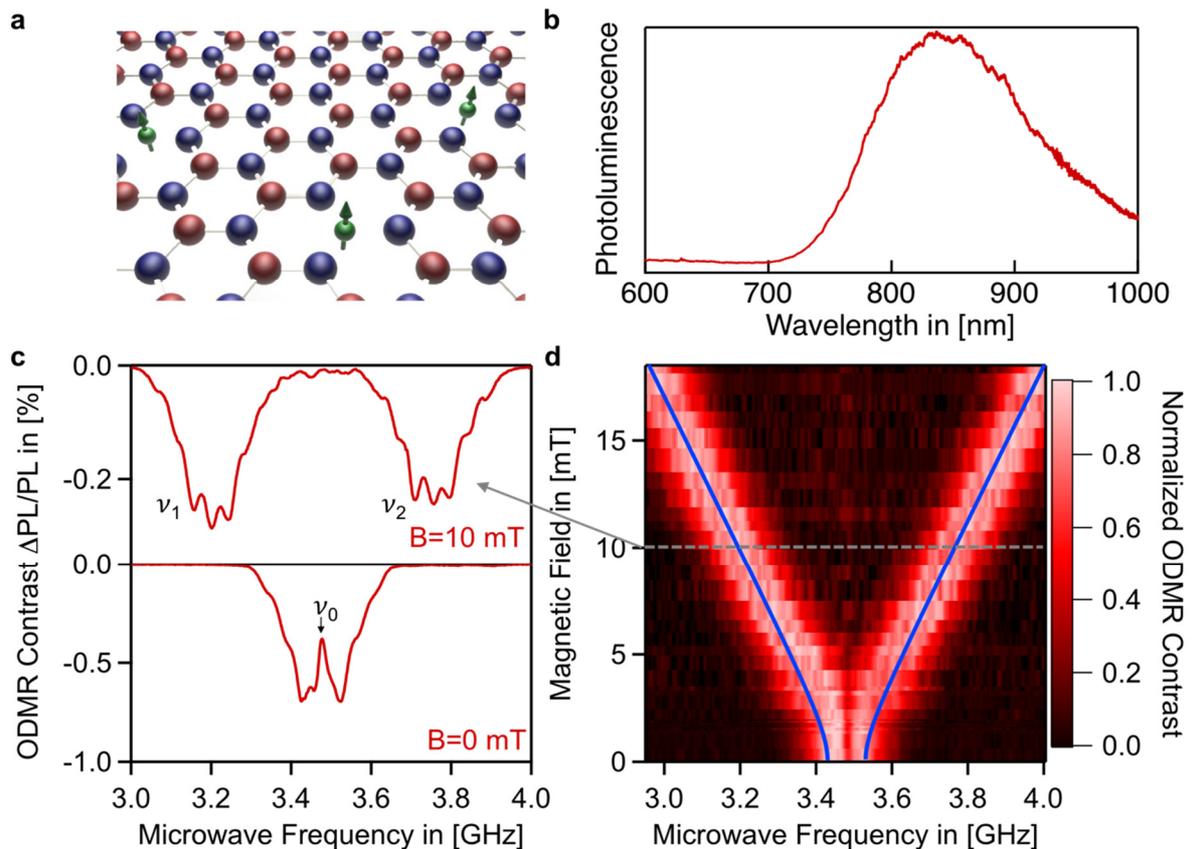

*Figure 1.* ODMR of an hBN single crystal at room temperature, T = 300 K. *a)* Schematic of an hBN monolayer and its crystalline hexagonal structure with alternating boron (red) and nitrogen (blue) atoms. The green arrows indicate the spins of the negatively-charged boron-vacancy defects, $V_B^-$. *b)* Photoluminescence spectrum of the sample at room temperature displaying a pronounced emission at 850 nm. *c)* ODMR spectra measured with zero magnetic field (bottom) and with magnetic field B = 10 mT (top); *d)* Dependence of ODMR frequencies $\nu_1$ and $\nu_2$ on the magnetic field (**B**∥**c**).

*Experimental (red) and fit (blue line) obtained using Equation 2 with parameters D/h = 3.48 GHz, E/h = 50 MHz and g=2.000.*

Most interestingly, we find that the PL from this hBN colour centre is spin-dependent. Figure 1c shows the ODMR spectrum recorded for an hBN single crystal at T = 300 K. In ODMR experiments, microwave-induced magnetic dipole transitions between spin-sublevels manifest as changes in PL intensity (ΔPL). The prerequisite for optical (PL) detection of EPR is thus the existence of a dependence between the optical excitation-recombination cycle and the defect's spin orientation. Figure 1c shows the spectrum of the investigated sample as normalized change of PL intensity (ΔPL/PL)—i.e. ODMR contrast—as a function of the applied microwave frequency $\nu$ for two static magnetic fields $B = 0$ and $B = 10\ mT$. Even without external magnetic field, the ODMR spectrum shows two distinct resonances $\nu_1$ and $\nu_2$, located symmetrically around the frequency $\nu_0$. We tentatively assign them to the $\Delta m_S = \pm 1$ spin transitions between triplet energy sublevels with completely lifted three-fold degeneracy, due to a splitting induced by dipolar interaction between the unpaired electron spins, forming the triplet. This so-called zero field splitting is described by parameters $D$ and $E$ which can be derived from the spectrum as $D/h = \nu_0$ and $\nu_{1,2} = (D \pm E)/h$. In order to verify this assignment, we have studied the dependence of the $\nu_1$ and $\nu_2$ resonant microwave frequencies on the magnitude of the external static magnetic field.

The evolution of the ODMR spectrum with the field applied parallel to the hexagonal *c*-axis (**B** ∥ **c**) of hBN is presented in Figure 1d. To explain the observed transitions and their variation with magnetic field, we use the standard spin Hamiltonian given by Equation 1 with *z* as the principle symmetry axis oriented perpendicular to the plane (collinear with the *c*-axis of the hBN crystal).

$$H = D(S_z^2 - S(S+1)/3) + E(S_x^2 - S_y^2) + g\mu_B \mathbf{B} \cdot \mathbf{S} \tag{1}$$

where *D* and *E* are the ZFS parameters, *S* = 1 is the total spin, *g* is the Landé factor, $\mu_B$ is the Bohr magneton, **B** is the static magnetic field and $S_{x,y,z}$ are the spin-1 operators. According to Equation 1 and for **B** applied parallel to the *c*-axis the resonant microwave frequencies at which the transitions occur vary as

$$\nu_{1,2} = \nu_0 \pm \frac{1}{h}\sqrt{E^2 + (g\mu_B B)^2} \tag{2}$$

where $\nu_0 = D/h$. The dependence of ODMR frequencies $\nu_1$ and $\nu_2$ on the magnetic field shown in Figure 1d can be perfectly fitted by Equation 2 with *g* = 2.000, |*D*|/h = 3.48 GHz and a small off-axial component of the ZFS *E*/h = 50 MHz. This demonstrates a highly symmetrical, almost uniaxial defect structure.

So far, we have shown that the investigated hBN defect is a *S* = 1 system, which can be optically addressed and read out using ODMR. However, from the ODMR measurements alone it is difficult to conclude whether this spin centre is in an excited, metastable or ground state, which is essential for determining the correct spin-dependent recombination pathway. In fact, we note that earlier results[26] proposed that defects in hBN had a spin-triplet metastable state, in contradiction to our current observations (see discussion below). A second consideration is that zero-field ODMR measurements on their own are not sufficient to deduce the microscopic structure of the defect. To address these points, we applied high-field ODMR and EPR to previously studied exfoliated hBN flakes[27] as well as to the hBN single crystal as studied by ODMR in Figure 1. We also note that the defects in the hBN single crystals were created in the same way as in the hBN-exfoliated flakes (see Methods).

Figure 2a shows EPR spectra taken at a fixed microwave frequency, $\nu$ = 9.4 GHz, while scanning the static magnetic field aligned parallel to the *c*-axis of the crystal (**B** ∥ **c**). Due to the amplitude modulation of the *B* field, EPR spectra usually look like first derivatives of the absorption signals. The spectra are recorded with (green trace) and without (black trace) optical excitation and consist of two groups of signals originating from two different types of paramagnetic species. The first group (centred at $B \approx 330$ mT) is characterised by $g = 2.003$ and consists of three EPR lines of nearly equal intensities corresponding to a paramagnetic centre with electron spin $S = ½$ interacting with a nuclear spin $I = 1$. The origin of such a splitting is hyperfine interaction. This group of lines remains the same with and without optical excitation and is not observed in ODMR under the same conditions (see Supplementary Figure S1). So, we conclude that spin transitions that cause these EPR signals are optically inactive. Although spin $S = ½$ centres were already reported in the late 1970s and assigned to a one-boron-centre (OBC) or three-boron-centre (TBC) defect[28], we believe that we are dealing with a $S = ½$ centre which occupies a boron site in the lattice that interacts with one $^{14}N$ ($I = 1$). However, this interpretation is beyond the scope of this work.

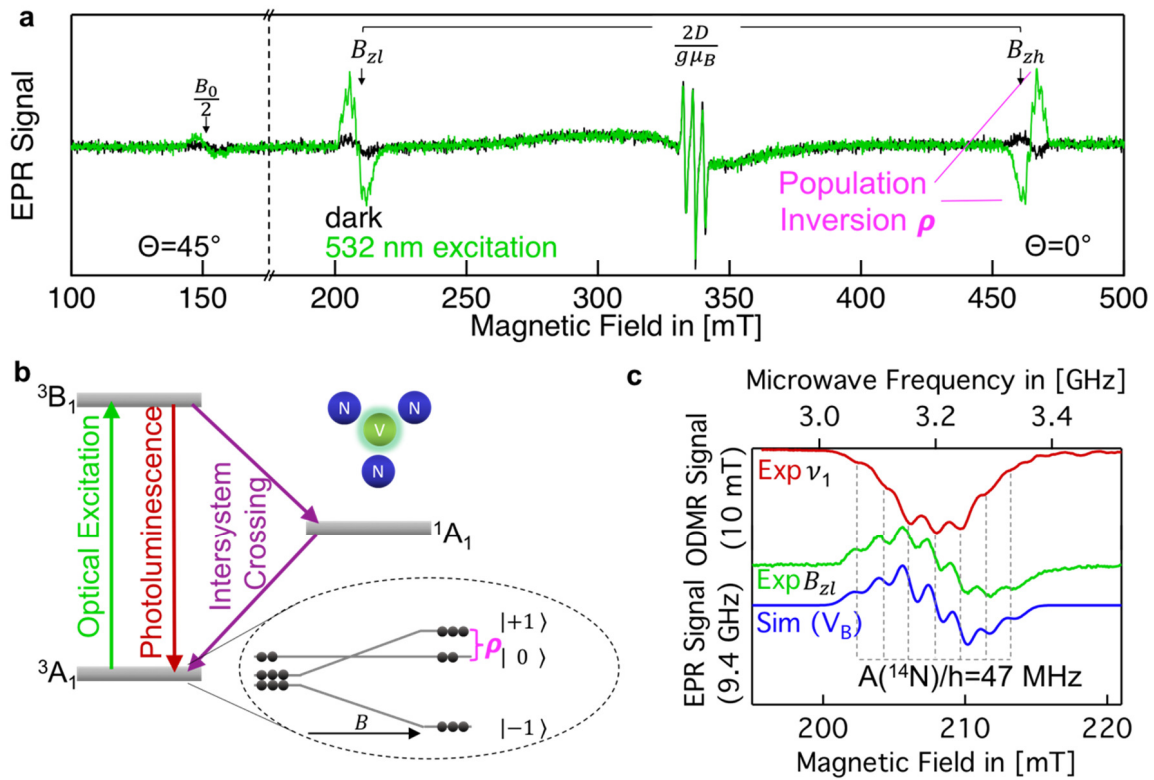

**Figure 2.** EPR studies of the $V_B^-$ centre in the hBN single crystal sample at $T = 5$ K. **a)** EPR spectra measured with (green trace) and without (black trace) 532 nm optical excitation in the magnetic field orientation **B ∥ c**. Doublet lines associated with the $S = 1$ centre with splitting of $\Delta B \approx 260$ mT (equivalent to $2|D|/h \approx 2 \times 3.6$ GHz) are labelled $B_{Zl}$, $B_{Zh}$. **b)** Simplified energy level diagram illustrating the optical spin polarization $\rho$ of the ground state with the optically pumped $m_S = \pm 1$ sublevels via intersystem crossing (ISC). **c)** Hyperfine splitting of EPR $B_{Zl}$ (green trace) and zero-field ODMR $\nu_1$ (red trace) lines. Vertical bars indicate seven transitions caused by hyperfine interaction with 3 equivalent nitrogen nuclei. The blue trace is the simulated EPR spectrum of the ($V_B^-$) defect with Eq.1, where the hyperfine term with a splitting constant $A/h = 47$ MHz is considered.

On the other hand, the second group of EPR lines in Figure 2a (labelled $B_{Zl}$, $B_{Zh}$) consists of two widely separated transitions, which are strongly responsive to optical excitation (532 nm laser). The splitting between the lines $B_{Zl}$ and $B_{Zh}$ ($\Delta B \approx 260$ mT) as well as the position of the $B_0/2$ signals at half-field are the same as in ODMR experiments under the same conditions performed on the hBN flakes (see Supplementary Figure S1). The EPR signals can be satisfyingly described by the spin-Hamiltonian (Equation 1) with the following Zeeman and ZFS parameters: $S = 1$, $g = 2.000$, $|D|/h = 3.6$ GHz. The ZFS parameter $E$, which became apparent in ODMR measurements at $B = 0$ (Figure 1c), could not be easily resolved in X-band EPR due to its small magnitude. It is however noticeable that the EPR measurements yield a slightly larger value for the ground-state ZFS parameter $|D|/h \approx 3.6$ GHz than what we determined from zero-field ODMR. This is expected for triplet spin centers[24] and we attribute this difference to a pronounced temperature dependence of $D$ varying between 3.6 GHz at $T = 5$ K (determined via EPR and ODMR) and 3.48 GHz at $T = 300$ K (ODMR). Assuming a linear dependence, this would lead to a slope of approximately –0.4 MHz/K, which can be reasonably expected for the temperature induced hBN lattice expansion.

Notably, the phases of the $B_{Zl}$ and $B_{Zh}$ signals shown in Fig. 2a become opposite (up-down and down-up) upon optical excitation and we observe emission rather than absorption for the EPR transition $B_{Zh}$. This can be explained by an optically-induced population inversion $\rho$ taking place amongst the spin-triplet sublevels of the ground state—with the $m_S = \pm 1$ sublevels, which are lower-lying than the $m_S = 0$ sublevel at $B = 0$. The order of the energy sublevels — with the $m_S = \pm 1$ being the lowest — is determined by the sign of the ZFS parameter $D$, which is negative in our case ($D < 0$). To prove this independently, we conducted EPR studies without optical pumping and found different signal intensities for $B_{Zl}$ and $B_{Zh}$ transitions in agreement with Boltzmann statistics (see Supplementary Figure S2).

In Figure 2b, we propose the tentative energy level scheme of the spin-defect consistent with our observations. Zeeman splitting of the $m_S = \pm 1$ levels results in the crossing of $m_S = +1$ and $m_S = 0$ sublevels, while optical pumping induces population transfer (via excited and metastable states) from $m_S = 0$ to $m_S = \pm 1$ and results in microwave emission at the $B_{Zh}$ field. We also want to emphasize that the EPR transitions between these triplet sublevels are also visible without optical excitation (black trace in Fig. 2a), suggesting that the triplet state we are looking at is in the ground state.

The combined EPR and ODMR data allow us to pinpoint the type of defect, although hBN can accommodate a large number of local defects in its lattice structure. These include common defects such as boron vacancies ($V_B$), nitrogen vacancies ($V_N$), anti-site complexes (e.g. a nitrogen atom on a boron site next to a vacancy ($N_B V_N$), or substitutional carbon-related defects like $C_B V_N$). The complex defects such as $V_N N_B$ and $V_N C_B$ have in-plane C$_{2v}$ symmetry, which is inconsistent with our observations of an almost axial defect with respect to the $c$-axis. On the other hand, the point defects $V_B$ and $V_N$ are characterized by the uniaxial D$_{3h}$ group symmetry with a C$_3$ rotation axis parallel to the $c$-axis. Both defects are thus compatible with our findings and must be considered. Recent theoretical investigations of point defects in hBN have shown that the high spin state ($S \geq 1$) is not a stable configuration for the $V_N$ defect, while the negatively-charged boron vacancy ($V_B^-$), with a $S = 1$ ground state, has been predicted to be stable[29, 30]. In addition, the optically-induced spin polarisation of the $m_S = \pm 1$ levels by intersystem crossing (ISC) has already been theoretically proposed for the $V_B^-$ triplet ground state[12]. Note, that we specify the negative charge for the $V_B^-$ defect, for the neutral $V_B$ does not possess a $S = 1$ ground state.

To unambiguously discern between the $V_N$ and $V_B^-$, we analysed the hyperfine structure of EPR and ODMR signals. Figure 2c shows the $B_{Zl}$ EPR line together with the zero-field ODMR $\nu_1$ transition. The hyperfine splitting is known to be due to the interaction of the electron spin with the surrounding nuclear spins, sometimes called superhyperfine interaction — which in turn reflects the nature of the nearest atoms (in our case either three boron or three nitrogen atoms). It can be described by adding the term $\sum_k S A_k I_k$ to Equation 1, where $A_k$ is the hyperfine interaction and $I_k$ is the nuclear spin. The number of observed hyperfine lines is seven (Fig. 3c). This is consistent with a spin $S = 1$ system interacting with $n = 3$ equivalent nitrogen atoms (nuclear spin $I = 1$, $^{14}$N, 99.63 % natural abundance) and it thus supports the case of the $V_B$ defect. Indeed, for $V_B$ there are $2nI + 1 = 7$ hyperfine transitions. Conversely, for $V_N$ there is hyperfine interaction of the electron spin with n = 3 equivalent boron atoms, each having two isotopes ($I = 3/2$, $^{11}$B, 80.2% natural abundance and $I = 3$, $^{10}$B, 19.8% natural abundance), resulting in $2nI + 1 = 10$ plus $2nI + 1 = 19$ transitions. The $V_N$ simulation shown in Supplementary Figure S3 takes the natural isotope abundance into account and is dominated by the 10 hyperfine transitions for $^{11}$B.

Consequently, the numerical simulations of the hyperfine structure show the best agreement for the boron vacancy model (see Figure 2c and Supplementary Figure S3) yielding a hyperfine splitting constant $A/h = 47$ MHz. Note that based on EPR data alone we cannot exclude an extrinsic defect with the same symmetry, such as e.g. $C_B^-$. To rule out this possibility, we performed irradiation of pristine hBN material (cf. Methods) with various species: ion-implantation with different ions (Li and Ga), as well as neutron irradiation. Regardless of the irradiation method, we observed the very same behaviour in all samples, hinting to the defect being indeed of intrinsic nature.

To test the symmetry of the defect more closely, we analysed the angular dependencies of the EPR signals for rotations around the polar and azimuthal angles $\theta$ and $\phi$, respectively. Figure 3 shows the EPR signals measured for a polar rotation ($\theta$) of the magnetic field from parallel (***B*** ∥ ***c***) to perpendicular (***B*** ⊥ ***c***) orientation and for azimuthal rotation ($\phi$) in the (0001) plane (***B*** ⊥ ***c***) of an hBN single crystal. The angular variation of the resonant magnetic field is described by numerical simulation of Eq. 1 employing the full set of the previously derived parameters (*S*, *g*, *D*, *E*). For both polar (Fig. 3a) and azimuthal rotation (Fig. 3b) we find exceptional good agreement of the overlaid simulated traces and the magnetic field positions of the experimentally observed transitions. We note that for the angular dependence measured in the plane of the σ bonds (***B***⊥***c*** in Fig. 3b), the splitting between the lines remains unchanged. These results point at the *c*-axis being the axis of symmetry of the almost uniaxial ODMR active triplet.

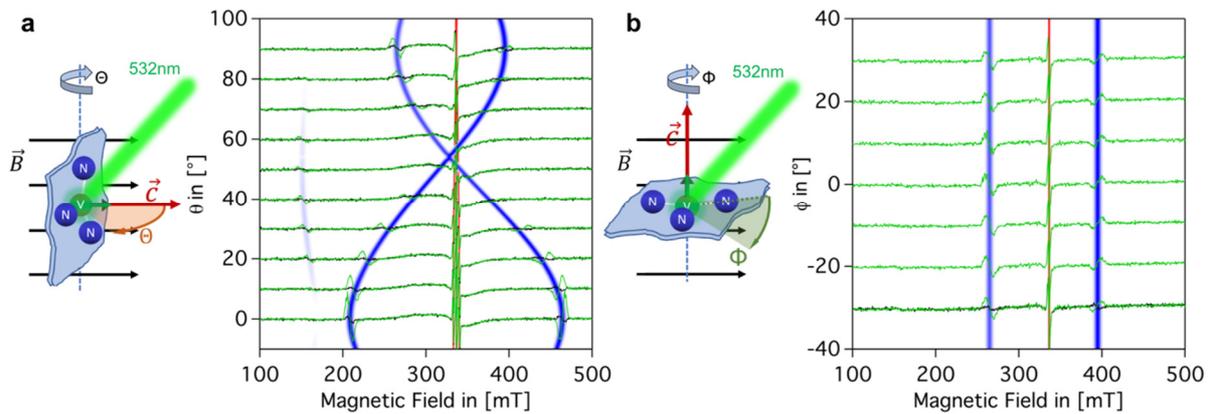

***Figure 3.*** *Angular dependence of the EPR spectra in an hBN single crystal. **(a)** Rotation of the magnetic field relative to the c-axis from parallel (**B** ∥ **c**) to perpendicular (**B** ⊥ **c**) orientation and **(b)** in the (0001) plane of hBN (**B** ⊥ **c**). Sketches explain the respective rotation schemes. Calculated angular dependencies are shown as blue traces. Angular variation of the S = ½ centre (central EPR group, see text) is shown in red. The detailed analysis of this region in shown in Supplementary Figure S4.*

To conclude, we have demonstrated room-temperature optical initialisation and readout of ensembles of intrinsic *S* = 1 colour centres in hBN. By rigorous EPR measurements performed on hBN single crystals and exfoliated flakes, we conclusively show that the investigated defect is the negatively-charged boron vacancy $V_B^-$. We demonstrated spin polarization of this spin defect under optical excitation and optically-induced population inversion in the triplet ground state, which provides the basis for coherent spin-manipulation. This stimulates further research in hBN-based heterostructures for quantum sensing applications and drives interest into deterministically engineering single $V_B^-$ centres. The work will also accelerate the research into the spin-optomechanics with hBN, particularly given the established theoretical framework[31] and advances in nanofabrication of resonators[32]. Further, with an increased control over isotopic purity, coherent manipulation of spin states may become feasible, yielding potentially high coherence times—as predicted theoretically[33]. In this context, the boron vacancy defect identified here may be advantageous due to weaker coupling of the defect electron spin with the surrounding $^{14}$N nuclear bath compared to other possible defect configurations.

**Acknowledgements**

V.D. acknowledges financial support from the DFG through the Würzburg-Dresden Cluster of Excellence on Complexity and Topology in Quantum Matter—ct.qmat (EXC 2147, project-id 39085490) and DY18/13-1. V.S. gratefully acknowledges the financial support of the Alexander von Humboldt (AvH) Foundation. The Australian Research council (via DP180100077, DP190101058 and DE180100810), the Asian Office of Aerospace Research and Development grant FA2386-17-1-4064, the Office of Naval Research Global under grant number N62909-18-1-2025 are gratefully acknowledged. I.A. is grateful for the Humboldt Foundation for their generous support.

**Methods**

**hBN samples**
The studied samples are single crystals and multilayered hBN flakes that were irradiated with various sources to create the defects. The same luminescence features were observed after neutron irradiation, as well as lithium or gallium ion implantation (see Supplementary Figure S6). Irradiation of the pristine hBN samples with different sources was conducted to test our hypothesis that the emitters are intrinsic in nature rather than due to inclusion of foreign atoms. Irrespective of the source, a strong luminescence emission appears in the near infrared when the irradiated samples are excited with a green laser, similar to the one presented in Figure 1b. All the EPR and ODMR measurements were carried out on the neutron irradiated samples. For more details about sample preparation by fast neutrons and thermal stability of produced defects see reference[27].

**Zero-field ODMR**
The zero-field ODMR measurements were performed with a confocal microscope setup. For optical excitation, a 532 nm laser was coupled into a 50 μm optical fibre and focussed onto the sample using a x10 objective (Olympus LMPLN10XIR), with the laser spot measuring approximately 10 μm in diameter. The laser power at the surface of the sample was 10 mW. The PL was then collected through the same objective and separated from the scattered laser light using a 650 nm short pass dichroic mirror and a 532 nm long pass filter. Behind the filter, it was coupled into a 600 μm optical fibre and detected using a Si avalanche photodiode (Thorlabs APD120A). The sample was placed on a 0.5 mm wide copper-stripline to apply the microwaves generated by a signal generator (Stanford Research Systems SG384) and amplified by an amplifier (Mini Circuits ZVE-3W-83+). ODMR was detected by a Signal Recovery 7230 lock-in amplifier referenced by on-off modulation of the microwaves. A permanent magnet was mounted below the sample to generate the external magnetic field. The precise magnetic field calibration was performed using a heterostructure, in which the hBN crystal was attached on top of a NV diamond sample and simultaneously optically excited using a confocal microscope (see Supplementary Figure S5), while the defects close to the interface were probed.

**EPR measurements**
EPR measurements were carried out with a modified Bruker spectrometer in the X-band regime (9.4 GHz) with a microwave power of 20 μW ($B_1 \approx 0.5$ μT). The hBN sample was placed inside an optically-accessible microwave cavity on a rotatable quartz rod in order to change the angle with respect to the external magnetic field. Using an Oxford cryostat, the sample was cooled down to 5 K. For optical excitation a 532 nm laser with 50 mW of power was directed through an optical window of the cavity. EPR spectral simulations were performed using EasySpin[34].

**High-field ODMR**
X-Band (9.4 GHz) ODMR measurements were performed in the same setup as for EPR using a second cavity window for optical readout. Transmitted light of the excitation wavelength is blocked by two filters (550 nm and 561 nm long pass) in order to detect the remaining broad PL of the emitting defects with a silicon photodiode (Hamamatsu S2281). The microwave source (Anritsu) is on-off modulated with

787 Hz and amplified to 2 W. The PL signal is preamplified (Femto DLCPA-200) and recorded by a lock-in amplifier (Signal Recovery 7230).